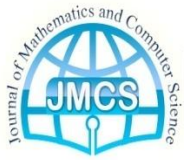
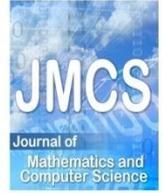

Contents list available at JMCS

# Journal of Mathematics and Computer Science

Journal Homepage: www.tjmcs.com

# Enhanced Slotted Aloha Mechanism by Introducing ZigZag Decoding

Abdellah ZAALOUL,   Abdelkrim HAQIQ
Computer, Networks, Mobility and Modeling laboratory
FST, Hassan 1st University, Settat, Morocco
e-NGN research group, Africa and Middle East

*zaaloul@gmail.com*, *ahaqiq@gmail.com*

*Abstract*

   Various random access mechanisms, such as Aloha protocol and its corresponding variants have been widely studied as efficient methods to coordinate the medium access among competing users. But when two or more wireless users transmit packets at the same time over the same channel a collisions occur. When this happens, the received packets are discarded and retransmissions are required, which is a waste of power and bandwidth. In such a situation one of the most important objectives is to find techniques to improve these protocols to reduce the number of collisions or to avoid them. Several studies have contributed to this problem.
In this paper, we propose a new approach named ZigZag decoding to enhance slotted Aloha mechanism by reducing the loss rate of packets colliding. We model the system by a Markov chain witch the number of backlogged packets is taken as the system state. We use a stochastic game to achieve our objective. We evaluate and compare the performances parameters of the proposed approach with those of slotted Aloha mechanism. All found results show that our approach is more efficient than the slotted Aloha mechanism.

## 1. Introduction

   Among components which characterize communication networks:  channels, and protocols. The protocols are the sets of rules and agreements among the communicating parties that dictate the behavior of the switches, and the channel is the physical medium over which signals, representing data, travel from one switch to another. Traditionally, networks use the point-to-point channels which are dedicated to a pair of users. The characteristic of these channels is that they generate no interference between a pair of nodes. The topology of Point-to-point channels, require be fixing, mostly determining at network design time. Subsequent topological changes are quite hard (and costly) to implement [1]. Over time, point-to-point channels are not economical, not available, or when dynamic topologies are required broadcast channels can be used. Naturally, broadcast channels appear in radio, satellite, and some local area networks. But, transmissions users over a broadcast channel interfere, and they collided. The major problem with multi-access is allocating the channel between the users, the node do not know when the other nodes have data to send.Which causes collision phenomenon, hence the necessity to coordinate transmissions. Thus, one of the major challenges in the design of multi-user wireless systems access is management of collision and handling interference, because the nature of wireless network is intrinsically different from the wired network when the medium is shared among several transmitters. The richest family of multiple access protocols is probably the Aloha family of protocols [2]. Aloha protocol is the first random access technique introduced. Some of these protocols are so simple that their implementation is straightforward.

Today many local area networks implement some sophisticated variants of this family's protocols. The success of a transmission when we use the Aloha family of protocols is not guaranteed in advance. The reason is that whenever two or more users are transmitting on the shared channel simultaneously, a collision occurs and the data cannot be received correctly, that involves the loss of all packets witch collided. Collision packets are known problem in wireless networks such as WLAN, 802.11, and 802.16 (WiMAX) [3, 4, 5, 6, 7, 9 ]. For example measurements from a production WLAN show that 10% of the sender-receiver pairs experience severe packet loss due to collisions [9]. Therefore, handling collision is one of the major challenges in the design of multi-user wireless network systems. Currently to limit and detect collisions, many protocols are used as CSMA slotted ALOHA [2]. These approaches are successful in many scenarios, but when it fails, as in the case of hidden terminals, the impact on the interfering senders is drastic; the senders either repeatedly collide and their throughputs plummet [4, 5, 9]. With slotted Aloha a collision occurs when two or more users are transmitting on the shared channel simultaneously. In order to reduce losses caused by collisions in a Slotted Aloha protocol, we propose, in this paper, to combine Slotted Aloha protocol with a technique named ZigZag decoding method [10]. It consists of reduction in the number of cases in which collisions occur, and then it can achieve throughput improvement in multi-user random access systems.

The study of the proposed method is based on Markov chain theory [26, 27] and stochastic game [25] which are strong and effective approaches to the analysis in telecommunication networks.

To evaluate performances of the proposed access method, we model the system by a stochastic game where the number of packets backlogged is taken as the system state. The Markov chain associated with this algorithm is then studied; its stationary distribution is determined, and the improvement of the average throughput and the average delay is then highlighted.

The rest of the paper is organized as follows: We begin by introducing a brief overview of ZigZag approach in section 2. In section 3 we give a brief overview related of work on random access schemes an ZigZag decoding approachs, in section 4 we proposed a model where we construct a Markov Model in a cooperative slotted Aloha combining with a ZigZag decoding where users maximize the total throughput of the system. In section 5 we studied the performance parameters. In section 6 we discuss the stability of system. In section 7 to numerically study and compare the properties of the proposed model with slotted Aloha. Section 6 Conclude the paper.

## 2. ZIG ZAG APPROACH OVERVIEW

Gollakota.S defines in [10] that ZigZag is a new decoding technique that increases random access methods resilience to collisions. A great advantage of ZigZag is that it requires no changes to the MAC layer and introduces no overheard in the case of no collision. If no collision occurs, ZigZag acts like a typical random access method. Another important aspect is that ZigZag achieves the same performance as if the colliding packets were a priori scheduled in separate time slots [10]. In ZigZag method, the receiver can decode two consecutive signals of two colliding packets and successfully receive both packets despite collision. In other words, if the same two packets collide twice, the receiver can receive both of those packets. Thus, the maximum achievable throughput of a wireless network can be significantly improved by using ZigZag decoding method. According to [10] there are some basic characteristics for ZigZag decoding method:

1- A ZigZag method can operate with unmodified network structure.
2- ZigZag method decreases the loss rate average.
3- Averaging over all sender-receiver pairs, including those that do not suffer from hidden terminals, the authors find that ZigZag improves the average throughput.

## 3. RELATED WORK

In this part of our work, we look some Prior works that has already analyzed this simple approach in many networks systems. Traditionally, a popular solution for wireless network is represented by a random multiple access. The most popular protocol employed and still employed is the slotted Aloha protocol [1, 4]. When multi-users send packets over common channel, random access schemes let to this population of users to share dynamically and opportunistically this channel. In practice, the level of coordination among the users wishing to access the channel is low or impossible (in many scenarios), this may be due to several reasons: for instance, to a lack of global information, to a too large user population size, or to the sporadic and unpredictable nature of users' access activity [13]. So, packets sent at the same time by several users fall in collision. Several works have studied and sought to resolve this collision phenomenon. The authors in [ 10] with ZigZag decoding show how to recover multiple collided packets in a 802.11 system when there are enough transmissions involving those packets. The main idea of Zig Zag approach is based on interference cancellation, the successive interference cancellation (SIC) technique was employed as a protocol in [16]. In this respect, SIC techniques has turned out to represent a major advance, allowing collisions be favorably exploited instead of being regarded as simply a waste.

Authors in [17] propose a scheme named CRDSA (Contention Resolution Diversity Slotted Aloha) exploiting SIC in the case of satellite access networks to remarkably improve the performance of the diversity slotted Aloha techniques (DSA) [18], this scheme consist of transmitting each packet twice in a medium access control (MAC) frame. The authors in [19] suggest ZigZag decoding, to extract the packets involved in the collisions. They present an algebraic representation of collisions and describe a general approach to recovering collisions using Analog Network Code ANC. They studied the effects of using ANC on the performance of MAC layers. The use of ZigZag without additional digital network coding has recently been considered by [20] to improve congestion control and maximize aggregate utility of the users.

In another related paper [21], the authors provide an abstraction of the multiple-access channel when ZigZag decoding is used at the receiver. They use this abstract model to analyze the delay and throughput performance of the system. Using various scenarios they conclude that the mean delivery time of the system with ZigZag decoding is strictly smaller than for a system with a centralized scheduler. Otherwise the stability region of the ZigZag decoding system is strictly greater.

## 4. PROPOSED MODEL AND PROBLEM FORMULATION

### 4.1. Proposed protocol

The Slotted aloha protocol is probably one of the most popular in the multiple access protocols family. It has long been used as random distributed medium access for radio channels. Indeed, it is so simple that its implementation is straightforward, and many local area networks of today implement some variants of Slotted aloha. In these protocols, packets are sent simultaneously by more than one user then they collide. Packets that are involved in a collision are backlogged and retransmitted later.

Original slotted-Aloha Protocol (Figure 1), is base on the following [2]:

- Time is divided into "slots" of one packet duration.
- When a node has a packet to send, it waits until the start of the next slot to send it.
- If no other nodes attempt transmission during that slot, the transmission is successful.
- Otherwise "collision", and packets involved in a collision are lost.
- Collided packets are retransmitted after a random delay.
- If a new packet arrives during a slot, it will be transmit in the next slot.
- If a transmission has a collision, node becomes backlogged.
- There are three immediate feedback states:  Idle (0), Success (1), Collision (C).

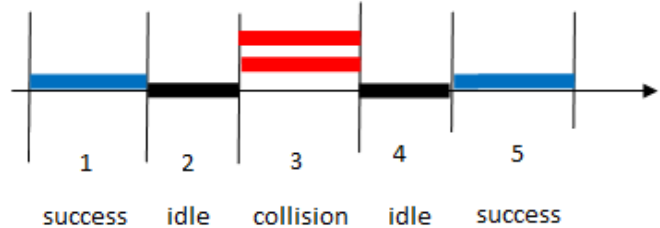

**Figure 1 : A timeline showing the various kinds of frames for slotted Aloha 'success', 'idle', 'collision'.**

Collision problem is state traditionally; when two or more packets are transmitted simultaneously a collision occurs. At the MAC layers, many solutions have been tested to eliminate collision hidden and exposed terminals [23].

A proposal is to alleviate the interference impact by learning the interference MAP, and taking scheduling decisions according to this MAP. At higher layers, network coding could also boost the system throughput, as demonstrated in [24].

ZigZag decoding is a new proposed approach for collision resolution [2]. In this approach if the same two packets collide twice, the receiver can receive both of those packets.

In the studied model we propose to combine the slotted Aloha medium access protocol with ZigZag decoding technique.

First the following assumptions are done:

- The frame size either one or two slots.
- At the beginning of a frame, all M users independently transmit (for the first time) or retransmit (in the case of a backlogged user) a packet.

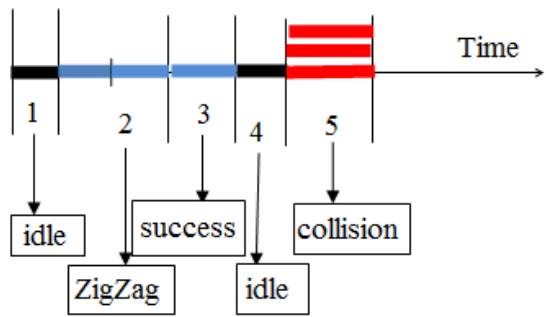

**Figure 2:  A timeline showing the various kinds of frames for proposed method: 'success', 'idle', 'collision' 'ZigZag'.**

In the proposed protocol, exactly one of the following four events happens (Figure 2):

1. Idle: nobody transmits any packet,
2. Success: exactly 1 user transmits a packet,
3. ZigZag: exactly 2 users transmit a packet,
4. Collision: when 3 or more users attempt transmission.

Then the receiver gives one of the 4 following feedback messages at the end of the first slot of the frame:

$$\text{Feedback} = \begin{cases} 0 & \text{if idle ( i, e, no packets attempted transmission)} \\ 1 & \text{if sucess (i, e, exactly 1 packet attempted transmission} \\ \text{ZigZag} & \text{if exactly 2 packets attempted transmission.} \\ C & \text{if 3 more packets attempted transmission} \end{cases}$$

With this scenario the frame has size 1 slot if the feedback is "0","1" or C, and if the feedback is "ZigZag" then the frame has a size of 2 slots. Therefore, using ZigZag decoding with Slotted Aloha, in our analysis we can redefine the term "collision" as follow: "A collision occurs on a slot when 3 or more users attempt transmission in a given time slot".

### 4.2. Problem formulation

In this subsection, we describe a slotted-Aloha MAC protocol combined with a ZigZag decoding technique and we construct a Markov Model based on [22], from which performance parameters are measured.

We consider a cellular system where M users transmit over a common channel to a base station.

We denote N the number of backlogged nodes (or equivalently, of backlogged packets) at the beginning of a slot.

It is obvious that N is a Markov Chain for which the state space is $E = \{0, \dots N\}$

For our model, we define and we adopt the following notations:

$Qr$: the vector of retransmission probability for all users (whose each entry is qr > 0),

$Qa$: the vector of transmission probability for all users (whose each entry is $p_a$).

Let $Q_a(i, N)$ be the probability that i unbacklogged nodes transmit packets in a given slot.

$$Q_a(i, N) = \binom{M-N}{i}(1 - p_a)^{M-N-i}(p_a). \tag{1}$$

And Let $Q_r(i, N)$ be the probability that i out of backlogged nodes retransmit packets in a given slot.

$$Q_r(i, N) = \binom{N}{i}(1 - q_r)^{N-i}(q_r)^i. \tag{2}$$

The transition probabilities of the Markov chain N are given by:

With

$$P_{N;N+i} = \begin{cases} Q_a(i,N) & 3 \leq i \leq M - N \\ Q_a(1,N)[1 - Q_r(0,N) - Q_r(1,N)] & i = 1 \\ Q_a(2,N)[1 - Q_r(0,N)] & i = 2 \\ Q_a(0,N)[1 - (Q_r(1,N) + Q_r(2,N))] + \cdots & \\ [Q_r(1,N) + Q_r(0,N)]Q_a(1,N) + Q_r(0,N)Q_a(2,N) & i = 0 \\ Q_a(0,N)Q_r(1,N) & i = -1 \\ Q_a(0,N)P_{ZIGZAG} & i = -2 \end{cases} \quad (3)$$

$$\text{And } P_{ZigZag} = \binom{2}{N}(1-qr)^{N-2}(qr)^2.$$

Since the state space is finite and all the states communicate between them the Markovian Chain N is ergodic.

Let $\pi(p_a, q_r)$ be the corresponding vector of steady state probabilities where its N th entry is $\pi_N(p_a, q_r)$ denotes the probability of N backlogged nodes.

This steady state distribution is solution of the following problem:

$$\begin{cases} \pi(p_a, q_r) = \pi(p_a, q_r) P(p_a, q_r) \\ \pi_N(p_a, q_r) \geq 0, N = 0 \ldots M \\ \sum_{N=0}^{M} \pi_N(p_a, q_r) = 1 \end{cases} \quad (4)$$

By computing recursively the steady state probabilities, we can obtain a solution to this problem, by calculating the performance metrics as in [2].

## 5. Performance metrics

### 5.1 Maximization of the Global Throughput:

The throughput of the system is defined as the sample average of the number of packets that are successfully transmitted; it is given almost surely by the constant:

$$Th(p_a q_r) = \sum_{N=1}^{M} \pi_N(p_a, q_r)[P_{N;N-1} + P_{N;N-2} + Q_a(0,N)Q_r(1,N) + Q_a(0,N)P_{ZIGZAG} + \pi_0(p_a, q_r)(Q_a(1,0)] = p_a \sum_{N=0}^{M} \pi_N(p_a, q_r)(M - N) \quad (5)$$

Therefore, we are interested to find an optimal solution of the following problem:

$$\max_{q_r} \; Th(p_a, q_r) \; s.t. \begin{cases} \pi(p_a, q_r) = \pi(p_a, q_r) P(p_a, q_r) \\ \pi_N(p_a, q_r) \geq 0, N = 0 \ldots M \\ \sum_{N=0}^{M} \pi_N(p_a, q_r) = 1 \end{cases} \quad (6)$$

We can also calculate the average number of backlogged packets by:

$$S_B(p_a, q_r) = \sum_{N=0}^{M} \pi_N(p_a, q_r) \cdot N \quad (7)$$

Using the formula $\sum_{N=0}^{M} \pi_N(p_a, q_r) = 1$ the throughput can be written as follow:

$$Th(p_a, q_r) = p_a(M - S_B). \quad (8)$$

### 5.2. Minimization of the Delay:

We can define the delay as the average time, in slots, that a packet takes from its source to the receiver. By little's formula [26], the delay is given by:

$$D(p_a, q_r) = \frac{Th(p_a, q_r) + S_B(P_a, q_r)}{Th(p_a, q_r)} = 1 + \frac{S_B(p_a, q_r)}{Th(p_a, q_r)} \quad (9)$$

The analysis of the equations (5) and (9) shows that maximizing the throughput is equivalent to minimizing the average delay of transmitted packets.

### 5.3. Performance measures for backlogged packets

An interesting alternative for measuring the performance of the system is to analyze the ability to serve packets awaiting retransmission. It has a great interest especially for real-time applications.

Let $T(p_a, q_r)$ is the average throughput of new packets arrived (crowned with success), so the average throughput for backlogged packets is given by: $\overline{T}(p_a, q_r) = Th(p_a, q_r) - T_{p_a, q_r}$.

Where $T(Pa, q_r)$ is calculated by:

$$T(p_a, q_r) = \sum_{N=0}^{M} \pi_N(p_a, q_r) Q_a(1, N) + Q_a(2, N))Q_r(0, N) \quad (10)$$

Thereafter we can calculate the expected delay $\overline{D}(p_a, q_r)$ for packets backlogged by applying little's formula [26]. Is given by:

$$\overline{D}(p_a, q_r) = 1 + \frac{S_B(p_a, q_r)}{\overline{T}(p_a, q_r)} \quad (11)$$

## 6. Stability

A qualitative approach to deal the performance of our protocol is to study its stability. Slotted Aloha is known to have a bi-stable behavior; it is natural to ask the following questions: 1) is our protocol suffers from the same problem as slotted Aloha? 2) If so, is there a difference between the two methods of access? 3) if not, under what conditions this variant would be better than slotted Aloha?

Let us define the drift $D_N$ in state N, as the expected change in backlog over one time slot given N backlogged packets. Thus, in [22], $D_N$ is the expected number of new arrivals accepted into the system from which we subs tract the expected number of successful transmissions in the slot.

The expected number of successful transmissions is just the probability of a successful transmission, defined as $P_{Succ}$. Thus,

$$D_N = (M - N)p_a - P_{Succ} \quad (12)$$

Where $P_{Succ}$ may be expressed as follows:

$$P_{Succ} = \big(Q_a(1, N) + Q_a(2, N)\big)Q_r(0, N) + \big(Q_r(1, N) + Q_r(2, N)\big)Q_a(0, N) \quad (13)$$

If $D_N < 0$, then the system tends to become less backlogged, and we are in a good situation.

Otherwise, if $D_N > 0$, then the system tends to become more backlogged, and the situation is not good.

## 7. Numerical result

We present in this section the numerical results that allow evaluating the performance metrics of system using the proposed method. On the other hand, we analyze and compare these metrics with those the slotted Aloha taken as reference.

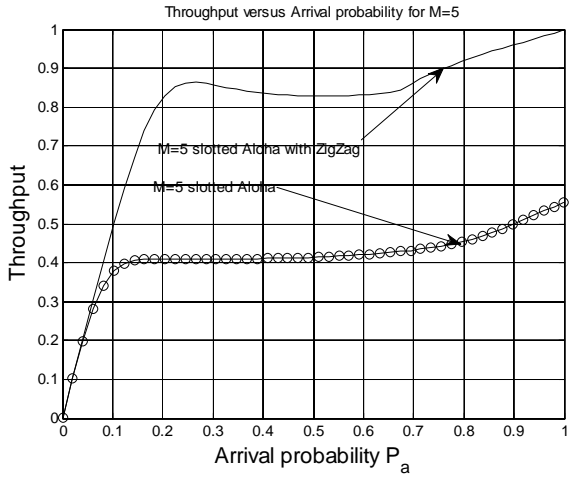 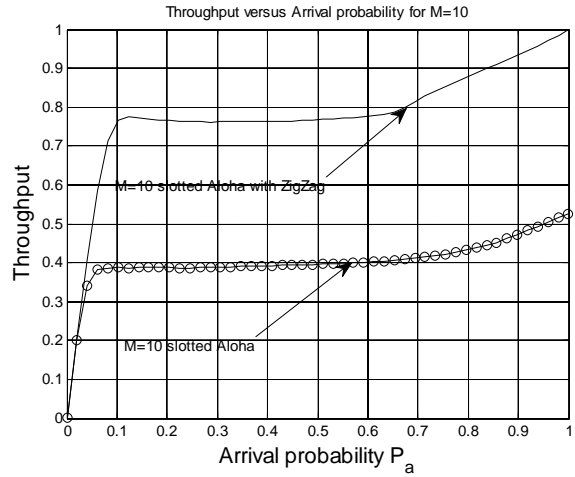

**Figure 3: throughput vs arrival probability for M=5.**    **Figure 4: throughput vs arrival probability for M=10.**

After solving equation (5) for M = 5 and M = 10. We observe that the average throughput, figure 3 and figure 4, has been significantly improved when using the slotted Aloha model with ZigZag decoding, either in low and high traffic; specially when the transmission probability becomes large ($p_a > 0.1$).

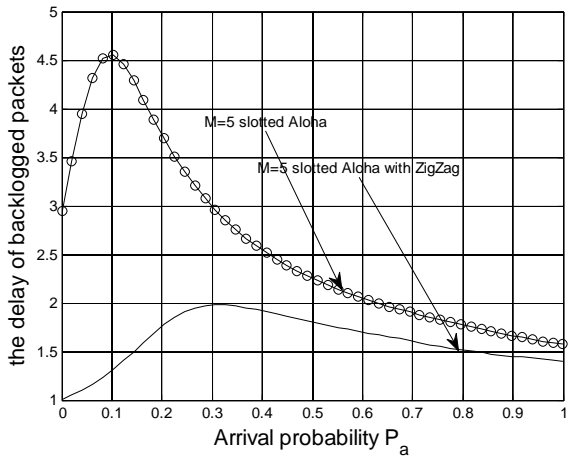 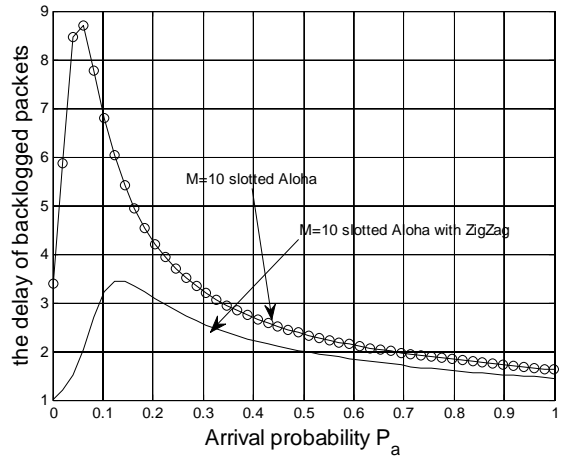

**Figure 5: the delay of backlogged packets vs arrival probability for M=5.**    **Figure 6: the delay of backlogged packets vs arrival Probability for M=10.**

The expected delay of backlogged packets is shown in figure 5 and figure 6. In all cases the proposed method significantly improves the delay of backlogged packets with reference to Slotted Aloha. This improvement is very important when the transmission probability isn't close to 1, and this is true either for heavy or low load.

However, when the load is heavy, we note that the improvement is very clear; the backlogged packets delay is almost reduced by 1/3, while the improvement is almost 1/2 at a low load.

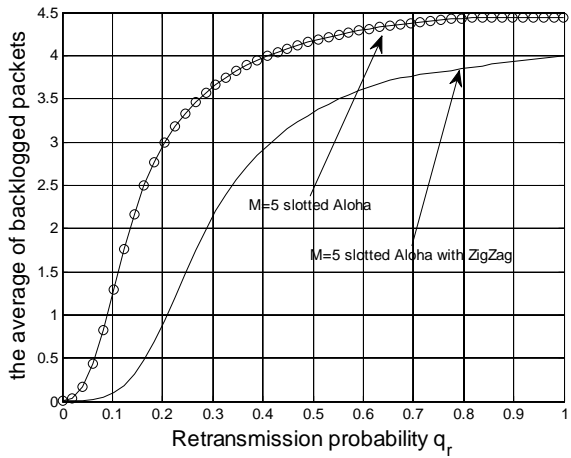 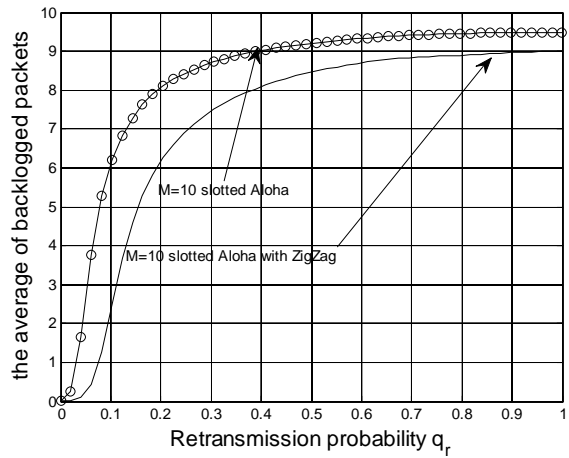

**Figure 7: the average of backlogged packets vs retransmission probability for M=10.**

**Figure 8: the average of backlogged packets vs retransmission probability for M=5.**

Figure 7 and figure 8 plot the average of backlogged packets versus the retransmission probability. It show that the new method reduce the number of this type of packets in the system. This improvement is more important when the system load is low.

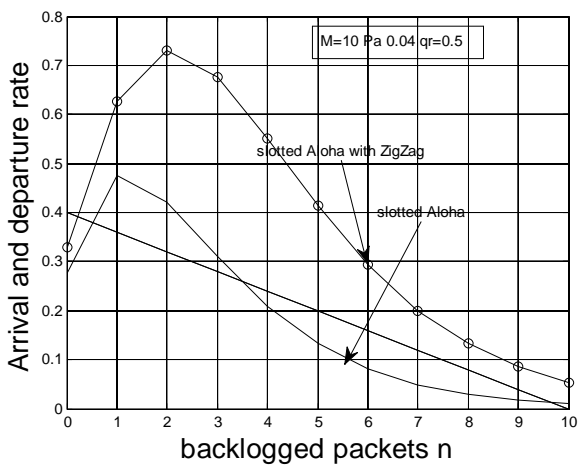 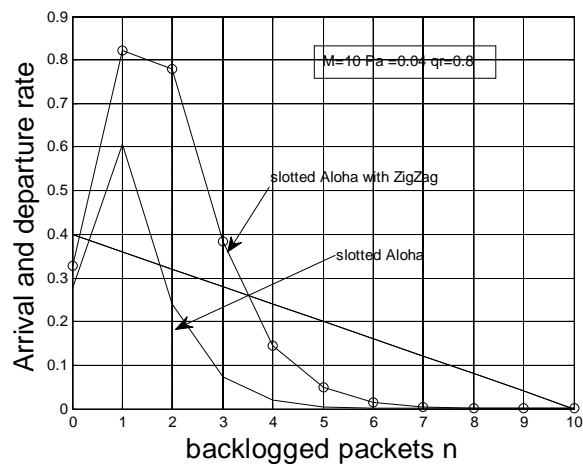

**Figure 9: Stability of slotted Aloha model with ZigZag Zig Zag and slotted Aloha for M=10 and $q_r = 0.8$.**

**Figure 10: Stability of slotted Aloha model with and slotted Aloha for M=10.**

The stability of the system is shown in figure 9 and figure 10 for a load M=10 users, and a transmission probability of $p_a = 0.04$ .

The stability conditions of the equilibrium strategy are based on the drift analysis of a Markov chain and the behavior of users when the number of backlogged users became very large. On the other hand, this version of slotted Aloha improved by ZigZag decoding in this paper assumes that users are heterogeneous in their bandwidth requirements and their neighboring user environment.

Figure 9 plotted with $q_r = 0.5$ show that our model is stable either low load or high load compared to slotted Aloha which has a bi-stable nature.

Figure 10 shows the impact of the retransmission probability $q_r$ (when $q_r$ increases, for $q_r = 0.8$ ) on the stability of this new approach. We affirm that our protocol is also bi-stable, but the bi-stability is delayed compared to Slotted Aloha, and the proposed method is more efficient in terms of probability of success. One way to remedy the problem of bi-stability would be to define a dynamic control of $q_r$ according to the instantaneous number of backlogged packets.

## 8. Conclusion

In this paper, a simple and practical extension of slotted Aloha improved by a new approach called ZigZag decoding is proposed. We constructed a Markov model and calculated its stationary distribution. This allowed us to calculate the various performance parameters such as throughput, expected delay and the average number of backlogged packets in the system. These performance indicators are used to study in a qualitative manner the stability of the protocol and quantitatively the gain in throughput and transmission delay. Our proposition is more efficient than slotted Aloha since it provides higher throughput and a low delay. Our solution is more efficient and could be capable of supporting real-time applications.